\documentclass[onecolumn,amsmath,amssymb,nofootinbib,aapt]{revtex4} 

\usepackage{amsmath}
\usepackage{graphicx}
\usepackage{graphics}
\usepackage{dcolumn}
\usepackage{bm}
\usepackage{hyperref}
\usepackage{xcolor}

\def\BEq{\begin{equation}}
\def\EEq{\end{equation}}
\def\BEqA{\begin{eqnarray}}
\def\EEqA{\end{eqnarray}}
\def\BEn{\begin{enumerate}}
\def\EEn{\end{enumerate}}
\def\BWT{\begin{widetext}}
\def\EWT{\end{widetext}}

\def\a{\alpha}

\def\b{\beta}

\begin{document}

\title{Globally stable, ghost-free hyperbolic square-root 
deformation of the Starobinsky model}

\author{Andrei Galiautdinov}
\email{ag1@uga.edu}
\affiliation{
Department of Physics and Astronomy, 
University of Georgia, Athens, Georgia 30602, USA
}
\begin{abstract}
We propose an exact, analytic deformation of the Starobinsky 
model governed by the strictly positive derivative of its 
Lagrangian, $f'(R) = \alpha R + \sqrt{\alpha^2 R^2 + 1}$, 
with $\alpha > 0$. This geometric hyperbolic square-root 
ansatz is designed to eliminate the well-known strong-coupling 
singularity that arises in quadratic $f(R)$ gravity when $f'(R)=0$. 
The construction seamlessly recovers general relativity at low 
curvatures and preserves the successful slow-roll inflationary 
plateau at extreme positive curvatures. In the limit $R \to -\infty$, 
the derivative $f'(R)$ asymptotes to zero strictly from above, 
removing the pathological branch associated with the vanishing 
of $f'(R)$. This guarantees that the only admissible 
constant-curvature ($R=A$) solutions correspond to standard 
Einstein spaces with an effective cosmological constant 
$\Lambda_{\text{eff}} \equiv A/4$, determined by the factorized 
field equations, 
$f'(A) \left( R_{\mu\nu} - \frac{A}{4} g_{\mu\nu} \right) = 0$. 
The first and second derivatives of the action, as well as the 
scalaron mass squared, remain strictly positive globally 
($f'>0$, $f''>0$, $m_s^2>0$ for all real $R$), ensuring a perfectly 
ghost-free and tachyon-free cosmological evolution across the 
entire spacetime manifold. In the Einstein frame, the dynamics of 
the scalaron is governed by the globally defined potential 
$V(\phi) = \frac{1}{8\alpha} [ 1 - (1 + 2\sqrt{2/3}\phi) \exp(-2\sqrt{2/3}\phi) ] 
+ \Lambda \exp(-2\sqrt{2/3}\phi)$, which naturally establishes 
an impenetrable energetic wall as $\phi \to -\infty$, offering a 
robust, globally stable mechanism for non-singular bouncing 
cosmologies. For $N = 60$ inflationary e-folds, the model predicts 
a scalar spectral index of $n_s \simeq 0.967$ and a strongly 
suppressed tensor-to-scalar ratio of $r \simeq 0.00083$, which 
position the proposed theory within the observationally 
favored parameter space of the Planck and BICEP/Keck Array 
baseline constraints.
\end{abstract}

\maketitle

\section{Introduction}

Modified theories of gravity, particularly $f(R)$ theories 
\cite{Capozziello2008, Capozziello2011, Nojiri2007, Nojiri2011, 
Nojiri2017, DeFelice2010, Sotiriou2010, Alvarez-Gaume2015}, 
are widely used to model the accelerated expansion phases of 
the universe. Among these, the Starobinsky model 
\cite{Starobinsky1980, Starobinsky1983, Vilenkin1985}, defined 
by the action $f(R) = R + \alpha R^2$ (where $\alpha = 1/(6M^2)$ 
and $M$ is the scalaron mass scale), is well-motivated by its 
renormalization properties, its natural suppression of higher-order 
quantum corrections, and its utility in inflationary cosmology. 
When transformed to its scalar-tensor equivalent in the Einstein 
frame, the $R^2$ term generates a flat potential plateau. This 
drives slow-roll inflation, yielding spectral indices consistent with 
modern cosmic microwave background (CMB) constraints from 
the Planck satellite \cite{Planck2018, BICEPKeck2021}.

In the metric formulation of $f(R)$ gravity, the gravitational 
field equations are given by,
\begin{equation}
\label{eq:f(R)-field-eqs}
f'(R) R_{\mu\nu}
- \frac{1}{2} f(R) g_{\mu\nu}
- \left[ \nabla_\mu \nabla_\nu - g_{\mu\nu} \Box \right] f'(R)
= 0, \quad \Box \equiv g^{\a\b}\nabla_{\a}\nabla_{\b},
\end{equation}
with constant-curvature solutions governed by the 
trace constraint,
\begin{equation}
\label{eq:f(A)-master-constraint}
A f'(A) - 2 f(A) = 0,
\end{equation}
where $R=A$ represents a constant spacetime curvature. 
Substituting Eq.~(\ref{eq:f(A)-master-constraint}) back into 
Eq.~(\ref{eq:f(R)-field-eqs}), we find that any 
constant-curvature metric must satisfy the factorized equation,
\begin{equation}
\label{eq:f(A)-field-eqs}
f'(A) \left( R_{\mu\nu} - \frac{A}{4} g_{\mu\nu} \right) = 0.
\end{equation}

In the quadratic $\Lambda$-Starobinsky model, 
\BEq
f(R) = R + \alpha R^2 - 2 \Lambda,
\EEq
the trace of the field equations fixes the macroscopic 
background to $A=4\Lambda$. Consequently, the reduced 
tensor equations acquire a multiplicative prefactor, 
$f'(A) = 1 + 8\alpha\Lambda$. In the specific parameter 
regime where this factor vanishes, the dynamical tensor 
constraints collapse identically, leaving only the scalar trace 
constraint $R=A$ to govern the geometry. As recently 
demonstrated \cite{Galiautdinov2026}, this resulting branch 
supports an arbitrary $1/r^2$ contribution in static 
spherically-symmetric solutions (cf., e.g., \cite{Dombriz2009, 
Sebastiani2011, Kehagias2015, Sebastiani2017, Calza2018, 
Golmoradifard2025}). This represents a mathematically 
degenerate configuration in which the effective gravitational 
coupling, $G_{\text{eff}} = G/f'(R)$, diverges, the graviton 
effectively decouples, and the associated scalar degree of 
freedom (the scalaron) loses its dynamical character 
(\cite{Dolgov2003, Faraoni2006, Amendola2007, Pogosian2008}).

This algebraic degeneracy points to a broader limitation of 
the generic quadratic formulation when extended to negative 
scalar curvatures, a regime that may be encountered during 
gravitational collapse or within a contracting cosmic phase. 
Crossing this critical boundary at the critical negative curvature 
$R_{\mathrm{crit}} = -1/(2\alpha)$ corresponding to
$f(R_{\mathrm{crit}})=0$, forces $f'(R) < 0$, which changes the sign 
of the kinetic term for the spin-2 graviton, turning it into 
a physically unacceptable ghost. Furthermore, the mass squared 
of the extra scalar degree of 
freedom in the Jordan frame is inversely proportional to the second 
derivative, $m_s^2 \simeq 1/(3f''(R))$. For the pure Starobinsky model, 
$f''(R) = 2\alpha$, providing a globally constant, strictly positive mass. 
However, the divergence at the $f'(R)=0$ boundary leads to 
multi-valuedness and a domain truncation in the Einstein frame. 

The necessity of structurally modifying $f(R)$ gravity to avoid 
pathological singularities has been widely discussed in the
literature. Well-known early approaches include the use 
of modifications with both positive and negative powers of 
curvature \cite{Nojiri2003} and the logarithmic modification 
with specifically constructed hyperbolic argument \cite{Appleby2010}, 
among others. In the context of late-time cosmic acceleration,
approaches such as the widely studied Hu-Sawicki model \cite{HuSawicki2007} 
have been highly successful. They utilize specific rational functional 
forms for $f(R)$ designed to mimic a cosmological constant at 
background curvatures while smoothly recovering exact general 
relativity in dense, high-curvature environments. This allows them 
to safely evade local solar system constraints via the chameleon 
mechanism \cite{Brax2008}. However, while such rational deformations 
are elegantly tailored for late-time dynamics, they are not naturally 
suited for the extreme ultraviolet regime of the early universe.

Additionally, other structures based on various square-root 
constructions have been explored to avoid singularities. 
For instance, Eddington-inspired Born-Infeld (EiBI) gravity 
uses a determinant-based square-root Lagrangian to regularize 
extreme densities \cite{Banados2010}, and specific 
phenomenological $f(R)$ models, such as $f(R) = \sqrt{R^2 - R_0^2}$, 
have been investigated to drive late-time cosmic acceleration \cite{Baghram2007}. 
However, these traditional approaches typically apply the square-root 
modification \emph{directly} at the level of Lagrangian density itself. 
This often complicate the task of preserving the required $f''(R) > 0$ stability 
condition across the entire curvature domain while simultaneously 
recovering the Starobinsky plateau.

To address these issues, we develop a specific modification of the 
$\Lambda$-Starobinsky model that preserves its inflationary phenomenology 
while mathematically forbidding $f'(R)$ from vanishing at any dynamically 
accessible curvature. By defining the theory through a 
\emph{hyperbolic square-root} deformation introduced at the level 
of the \textit{derivative}, $f'(R)$, we construct a globally well-behaved 
geometry. We demonstrate that this model maintains the Starobinsky plateau 
for $R>0$, recovers general relativity at $R=0$, and avoids the pathological 
$f'(R)=0$ crossing entirely. This formulation resolves the negative-curvature singularity 
and ensures both classical and perturbative stability ($f''(R) > 0$) across the 
spacetime manifold for \emph{all} curvatures $R \in (-\infty, +\infty)$.

\section{The hyperbolic square-root deformation model}

To construct a globally stable theory that evades the $f'(R)=0$ degeneracy 
while preserving the $R^2$ inflationary regime, we require a derivative 
function $f'(R)$ that is strictly positive, monotonically increasing (to ensure 
$f''(R) > 0$), and asymptotes to linear growth at large positive $R$. 

\begin{figure}[!ht]
\centering
\includegraphics[angle=0,width=0.55\linewidth]{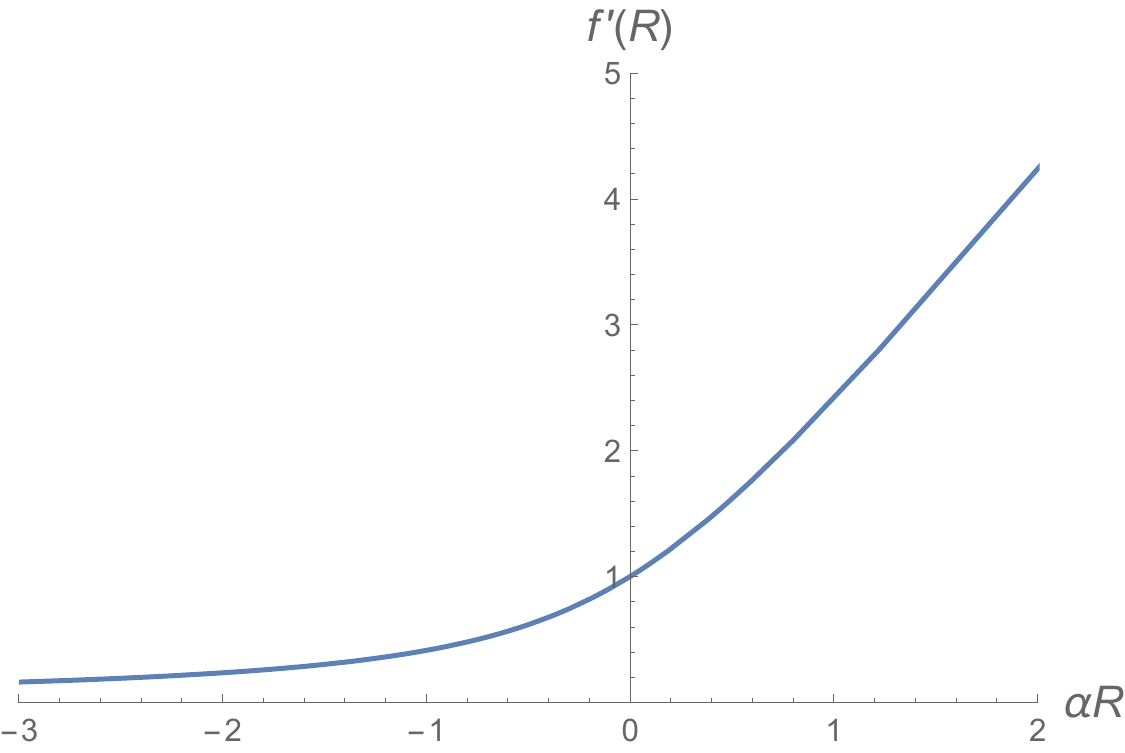}
\caption{\label{fig:01} The first derivative of the modified action, 
$f'(R) = \alpha R + \sqrt{\alpha^2 R^2 + 1}$, which serves as the 
foundational geometric ansatz for our deformed Starobinsky model. 
The function $f'(R)$ remains strictly positive for all real curvature 
values $R \in (-\infty, +\infty)$. This positivity ensures that the effective 
gravitational coupling $G_{\mathrm{eff}} = G/f'(R)$ is finite and positive, 
preventing the spin-2 graviton from becoming a ghost and ensuring 
the conformal mapping to the Einstein frame remains non-singular.}
\end{figure}

\begin{figure}[!ht]
\centering
\includegraphics[angle=0,width=0.55\linewidth]{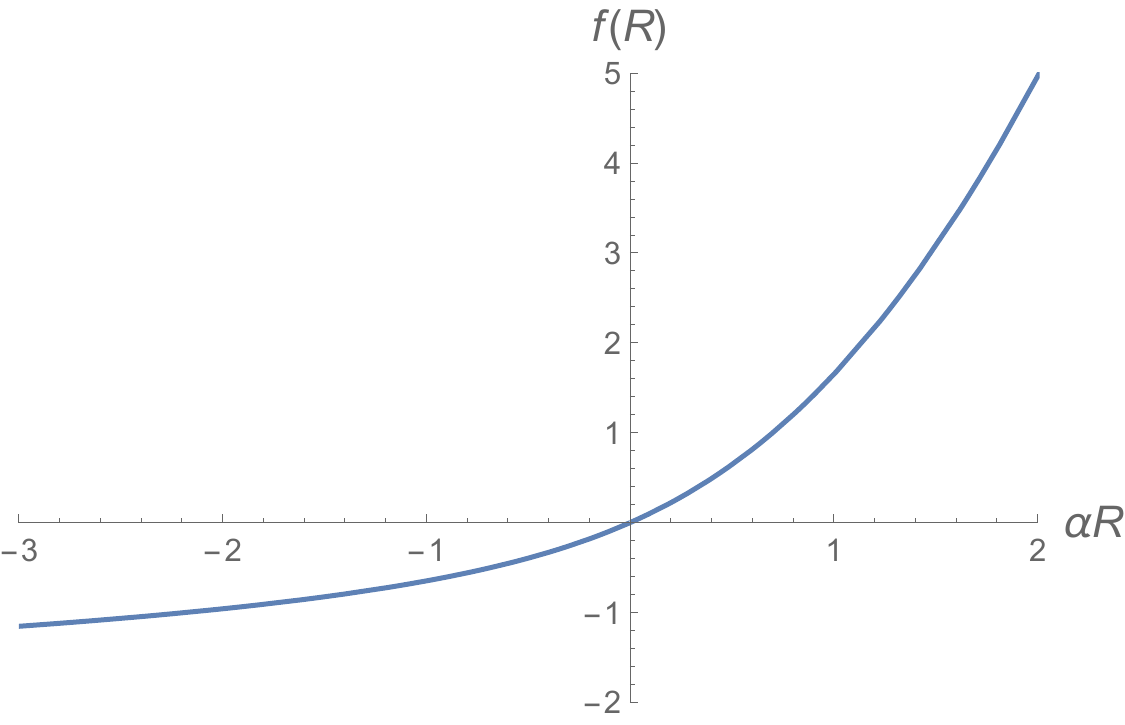}
\caption{\label{fig:02} The Lagrangian density $f(R)$ for the deformed 
Starobinsky model, evaluated with a vanishing cosmological constant 
($\Lambda=0$). Obtained by integrating the derivative shown in 
Fig.\ \ref{fig:01}, this function recovers the linear Einstein-Hilbert limit 
($f(R) \simeq R$) at low curvatures, while transitioning to the 
$\mathcal{O}(R^2)$ scaling required for slow-roll inflation at large 
positive curvatures.}
\end{figure}

We propose the following functional form for the \emph{derivative} of 
the modified gravity action,
\begin{equation}
\label{eq:fprime}
f'(R) = \alpha R + \sqrt{\alpha^2 R^2 + 1},
\end{equation}
shown in Fig.\ \ref{fig:01}, where $\alpha > 0$ is the characteristic 
coupling parameter with dimensions of inverse mass squared. Integrating 
Eq.~(\ref{eq:fprime}) with respect to the Ricci scalar yields the exact analytic 
form of the Lagrangian (Fig.\ \ref{fig:02}),
\begin{equation}
\label{eq:fR} 
f(R) = \frac{1}{2}R \left(\alpha R + \sqrt{\alpha^2 R^2 + 1}\right) 
+ \frac{1}{2\alpha} \mathrm{arcsinh}(\alpha R) - 2 \Lambda .
\end{equation}
This function constitutes the foundation of the proposed model. 
We now systematically analyze its behavior across the three critical 
curvature regimes.

For small curvatures where $|\alpha R| \ll 1$, we can Taylor expand 
Eq.~(\ref{eq:fR}) around $R=0$,
\begin{equation}
f(R) \simeq R - 2 \Lambda + \left( \frac{1}{2}\alpha R^2 
    + \frac{1}{6}\alpha^2 R^3 + \mathcal{O}(R^5) \right),
\end{equation}
smoothly recovering general relativity and satisfying local solar 
system constraints.

In the high-curvature regime ($R \to +\infty$), the term inside 
the square root is dominated by $\alpha^2 R^2$. The derivative 
simplifies to $f'(R) \simeq 2\alpha R$. Consequently, the action 
asymptotes to
\begin{equation}
f(R) \simeq \alpha R^2,
\end{equation}
confirming that this deformation preserves the $R^2$ dominance 
required at high positive curvatures. When mapped to the Einstein 
frame, this behavior supports the flat potential plateau necessary 
for viable slow-roll inflation.

In the deep negative curvature regime ($R \to -\infty$), 
$\sqrt{\alpha^2 R^2 + 1} \simeq |\alpha R| = -\alpha R$. 
Expanding the square root strictly, we find
\begin{equation}
\sqrt{\alpha^2 R^2 + 1} \simeq -\alpha R - \frac{1}{2\alpha R}.
\end{equation}
Substituting this into Eq.~(\ref{eq:fprime}) yields the asymptotic 
behavior of the derivative,
\begin{equation}
f'(R) \simeq -\frac{1}{2\alpha R}.
\end{equation}
Because $R$ is strictly negative in this regime, $f'(R)$ approaches 
zero asymptotically from the positive side ($f'(R) \to 0^+$). Since 
the derivative never crosses zero, the strong-coupling singularity 
of standard Starobinsky gravity is avoided. 

\begin{figure}[!ht]
\centering
\includegraphics[angle=0,width=0.55\linewidth]{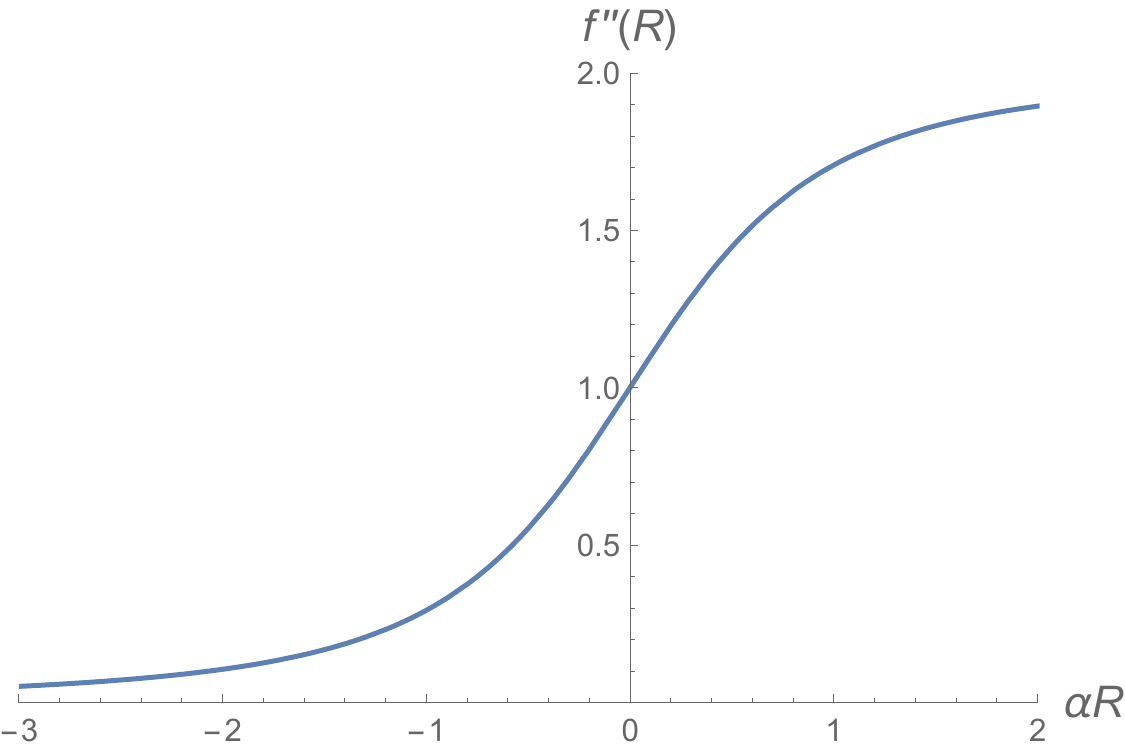}
\caption{\label{fig:03} The second derivative of the modified action, 
$f''(R)$, plotted as a function of the background Ricci scalar for our 
exact square-root deformation. The strict positivity $f''(R) > 0$ is 
maintained across the entire curvature domain $R \in (-\infty, +\infty)$. 
This is a necessary condition in $f(R)$ gravity to avoid the 
Dolgov-Kawasaki instability and ensures the conformal mapping to 
the Einstein frame remains non-singular.}
\end{figure}

For the theory to be free of ghosts and tachyonic instabilities, it must 
satisfy $f'(R) > 0$ and $f''(R) > 0$ globally. From Eq.~(\ref{eq:fprime}), 
it is evident that $f'(R) > 0$ for all $R \in \mathbb{R}$. Taking the second 
derivative yields
\begin{equation}
\label{eq:fdoubleprime}
f''(R) 
= \alpha + \frac{\alpha^2 R}{\sqrt{\alpha^2 R^2 + 1}} 
= \frac{\alpha f'(R)}{\sqrt{\alpha^2 R^2 + 1}}.
\end{equation}
Since $f'(R)$ and the square root are both strictly positive, $f''(R) > 0$ 
everywhere. This condition ensures the absence of the Dolgov-Kawasaki 
instability (Fig.\ \ref{fig:03}).

Finally, we calculate the exact mass squared of the Jordan-frame 
scalaron. Using the standard formula,
\begin{equation}
m_s^2 = \frac{1}{3}\left(\frac{f'(R)}{f''(R)} - R\right),
\end{equation} 
and Eq.~(\ref{eq:fdoubleprime}), we find a simple algebraic expression 
for the scalaron mass,
\begin{equation}
m_s^2 = \frac{1}{3} \left( \frac{1}{\alpha}\sqrt{\alpha^2 R^2 + 1} - R \right).
\end{equation}
Because $\sqrt{\alpha^2 R^2 + 1} > \sqrt{\alpha^2 R^2} 
= |\alpha R| \geq \alpha R$, it follows that $m_s^2$ is strictly 
positive across the entire curvature domain. The vacuum is stable, 
possessing no tachyonic regimes.

\begin{figure}[!ht]
\centering
\includegraphics[angle=0,width=0.55\linewidth]{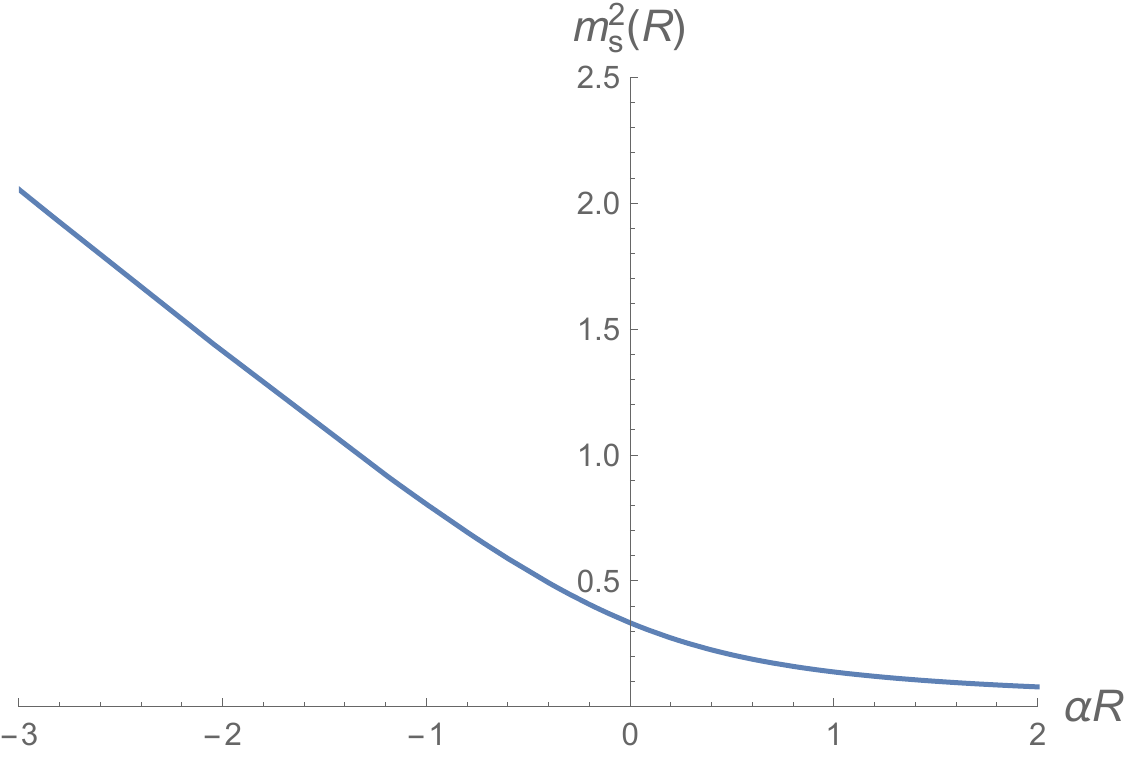}
\caption{\label{fig:04} The effective scalaron mass squared, 
$m_{\mathrm{s}}^2$, plotted as a continuous function of the 
background Ricci scalar $R$ for the square-root deformed model. 
Importantly, the mass squared remains strictly positive 
($m_{\mathrm{s}}^2 > 0$) and finite across the entire global 
curvature domain $R \in (-\infty, +\infty)$. This positivity ensures 
the classical and perturbative stability of the spacetime, preventing 
the scalar degree of freedom from becoming tachyonic.}
\end{figure}

\section{Static Spherically Symmetric Solutions and the Absence of Exotic Hair}
\label{sec:spherical_solutions}

Having established the structural properties and perturbative 
stability of the hyperbolic square-root deformation, we now 
examine its macroscopic astrophysical regime. A primary physical 
motivation for introducing this model was the elimination of the strongly 
coupled sector found in the standard quadratic theory. We demonstrate 
this by considering the static, spherically symmetric solutions of the model 
and confirming the uniqueness of the standard general relativistic black 
hole spacetimes.

We adopt the simplest ansatz for a static, spherically 
symmetric line element,
\begin{equation}
ds^2 = -h(r) dt^2 + \frac{dr^2}{h(r)} + r^2 d\Omega^2,
\end{equation}
where $d\Omega^2 = d\theta^2 + \sin^2\theta d\phi^2$ is the 
metric on the unit two-sphere. The scalar curvature for this geometry 
is given by
\begin{equation}
R = -\left(h''(r) + \frac{2 \big[ 2 r h'(r) + h(r) - 1 \big]}{r^2}\right).
\end{equation}

Restricting our attention to the dynamically dominant family of 
constant-curvature solutions, where $R = A = \text{constant}$, the 
fourth-order field equations reduce to the factorized form given 
in Eq.~(\ref{eq:f(A)-field-eqs}). 
Because our deformation guarantees that $f'(A) > 0$ for all real 
curvatures, this multiplicative prefactor never vanishes. Consequently, 
the equations simplify directly to the standard vacuum, 
Einstein equations,
\begin{equation}
R_{\mu\nu} - \frac{A}{4} g_{\mu\nu} = 0.
\end{equation}
This geometry characterizes an Einstein space endowed with an 
effective macroscopic cosmological constant,
\begin{equation}
\Lambda_{\text{eff}} = \frac{A}{4},
\end{equation}
governed by the universal master constraint, 
Eq.\ (\ref{eq:f(A)-master-constraint}). 
Solving the reduced Einstein equations yields the unique macroscopic 
spacetime, the standard Schwarzschild--(anti)de Sitter metric,
\begin{equation}
h(r) = 1 - \frac{2 M}{r} - \frac{\Lambda_{\text{eff}}}{3} r^2.
\end{equation}

This uniqueness contrasts sharply with the behavior of the standard 
quadratic $\Lambda$-Starobinsky theory. In the quadratic model, 
a secondary family of solutions emerges precisely when 
$f'(A) = 1 + 2\alpha A = 0$. On this degenerate branch, the tensor 
equations are trivially satisfied, permitting an arbitrary $1/r^2$ 
integration constant in the metric \cite{Galiautdinov2026},
\begin{equation}
h(r)_{\rm degenerate} 
= 
1 - \frac{2 M}{r} + \frac{c_2}{r^2} - \frac{\Lambda_{\text{eff}}}{3} r^2.
\end{equation}
This $c_2/r^2$ term represents ``exotic geometric hair.'' It arises 
strictly as an integration artifact of the $f'(A)=0$ sector. This degenerate 
branch simultaneously suffers from a divergent effective gravitational 
coupling, identically vanishing Noether charges, and ill-defined horizon 
thermodynamics.

Because our proposed deformation enforces $f'(R) > 0$ globally, 
the $f'(R)=0$ condition is never realized. This leads to four direct 
physical consequences governing the macroscopic phenomenology 
of the model:
\begin{enumerate}
\item \textbf{Absence of Exotic Hair:} The static, spherically symmetric 
metric must necessarily satisfy the standard Einstein equations with 
an effective cosmological constant. The appearance of an arbitrary 
$1/r^2$ geometric term is dynamically forbidden.
\item \textbf{Uniqueness of the SdS Spacetime:} The standard 
Schwarzschild--(anti)de Sitter solution is the unique $\Lambda$-vacuum 
geometry within the static, spherically symmetric, constant-curvature sector, 
parameterized solely by the physical ADM mass $M$ and the effective 
vacuum energy.
\item \textbf{Well-Defined Thermodynamics:} Macroscopic 
thermodynamic quantities remain finite and well-behaved. The Wald 
entropy (\cite{Wald1993, Vivek1994}) of the black hole horizon ($r_H$), given by 
$S = \frac{{\cal A}_H}{4 G} f'(A_H)$, where ${\cal A}_H$ is the horizon 
area, is strictly positive. The finite macroscopic horizon curvature 
guarantees that the effective gravitational coupling at the horizon 
remains perturbative.
\item \textbf{Linear Stability:} Scalar geometric perturbations 
are governed by the positive scalaron mass 
squared ($m_{\text{s}}^2 > 0$), ensuring the classical linear stability 
of the macroscopic black hole solution against spontaneous geometric 
collapse.
\end{enumerate}

The uniqueness of the Schwarzschild--(anti)de Sitter 
solution ensures that this modification 
preserves the standard strong-field black hole structure of general 
relativity while excising the problematic strong-coupling branch. 
Consequently, test particle dynamics, gravitational lensing, and 
horizon thermodynamics remain unambiguous and free from 
unphysical metric artifacts.

\section{Einstein Frame Dynamics and the Potential Wall}

To fully comprehend the physical implications of this deformation, 
particularly concerning cosmic inflation and the resolution of the 
negative-curvature singularity, it is instructive to analyze the theory in 
its scalar-tensor form. By applying a conformal transformation, we can 
map the Jordan-frame $f(R)$ action into the Einstein frame, wherein 
the modified gravitational dynamics is recast as standard general relativity 
coupled to a canonical scalar field $\phi$, the scalaron.

\subsection{Conformal Transformation and the Exact Mapping}

The conformal factor linking the two frames is defined by the derivative 
of the action,
\begin{equation}
\Omega^2 = f'(R) = \alpha R + \sqrt{\alpha^2 R^2 + 1}.
\end{equation}
The canonical scalar field $\phi$ is given by the standard relation,
\begin{equation}
\kappa \phi = \sqrt{\frac{3}{2}} \ln f'(R),
\end{equation}
where $\kappa = \sqrt{8\pi G} = 1/M_{\mathrm{Pl}}$. For mathematical 
brevity in the following derivations, we set $M_{\mathrm{Pl}} = 1$ 
and define the dimensionless field variable $x \equiv \sqrt{2/3} \, \phi$, 
yielding the direct relation $f'(R) = e^x$.

By setting 
\begin{equation}
f'(R) = e^x = \alpha R + \sqrt{\alpha^2 R^2 + 1},
\end{equation} 
and using the identity
\begin{equation}
\label{eq:arcsinhTOln}
\mathrm{arcsinh}(\alpha R) 
    = \ln\left(\alpha R + \sqrt{\alpha^2 R^2 + 1}\right),
\end{equation}
we can analytically invert the relationship to express the Jordan-frame 
Ricci scalar as an exact function of the scalar field,
\begin{equation}
\label{eq:Rphi}
\alpha R = \sinh(x) = \sinh\left(\sqrt{\frac{2}{3}}\phi\right).
\end{equation}
This perfectly smooth, global, and single-valued algebraic mapping 
is the mathematical mechanism that prevents the formation of any 
geometric cusps or multivalued potentials.

\subsection{The Exact Scalar Potential}

\begin{figure}[!ht]
\centering
\includegraphics[angle=0,width=0.6\linewidth]{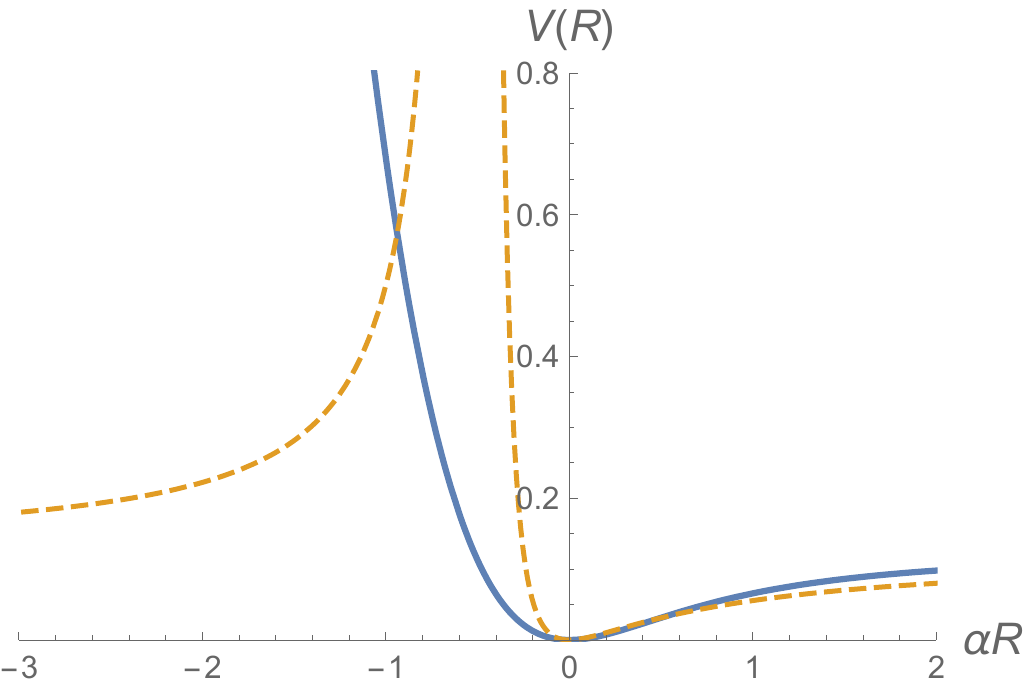}
\caption{\label{fig:05} The scalaron potential expressed as a function 
of the Jordan-frame Ricci scalar, $V(R)$, evaluated at $\Lambda = 0$ 
for our exact square-root deformation (solid blue) and the original 
Starobinsky model (dashed orange). Both potentials exhibit a stable 
Minkowski vacuum at $R=0$ and behave identically in the high-curvature 
inflationary regime ($R \to +\infty$). However, their dynamics sharply 
diverge at negative curvatures. The standard Starobinsky potential suffers 
a pathological pole at $R_{\mathrm{crit}}=-1/(2\alpha)$, where $f'(R)=0$. 
At this boundary, the effective gravitational coupling diverges, and the 
conformal mapping becomes singular. In contrast, the deformed model's 
potential smoothly bypasses this singularity, remaining finite and globally 
well-defined for all $R \in (-\infty, +\infty)$.}
\end{figure}

The dynamics of the scalar field in the Einstein frame is described 
by the effective potential $V(\phi)$, defined universally for $f(R)$ 
theories as
\begin{equation}
\label{eq:Vdef}
V(\phi) = \frac{R f'(R) - f(R)}{2 f'(R)^2},
\end{equation}
which in our model becomes
\begin{equation}
V(R) = \frac{ \alpha R \left(\sqrt{1+\alpha^2 R^2}+\alpha R \right)
+4 \alpha \Lambda - \mathrm{arcsinh}(\alpha R)}
{4 \alpha  \left(\sqrt{1+\alpha^2 R^2}+\alpha R \right)^2},
\end{equation}
as depicted in Fig.\ \ref{fig:05}.

\begin{figure}[t]
\centering
\includegraphics[angle=0,width=0.6\linewidth]{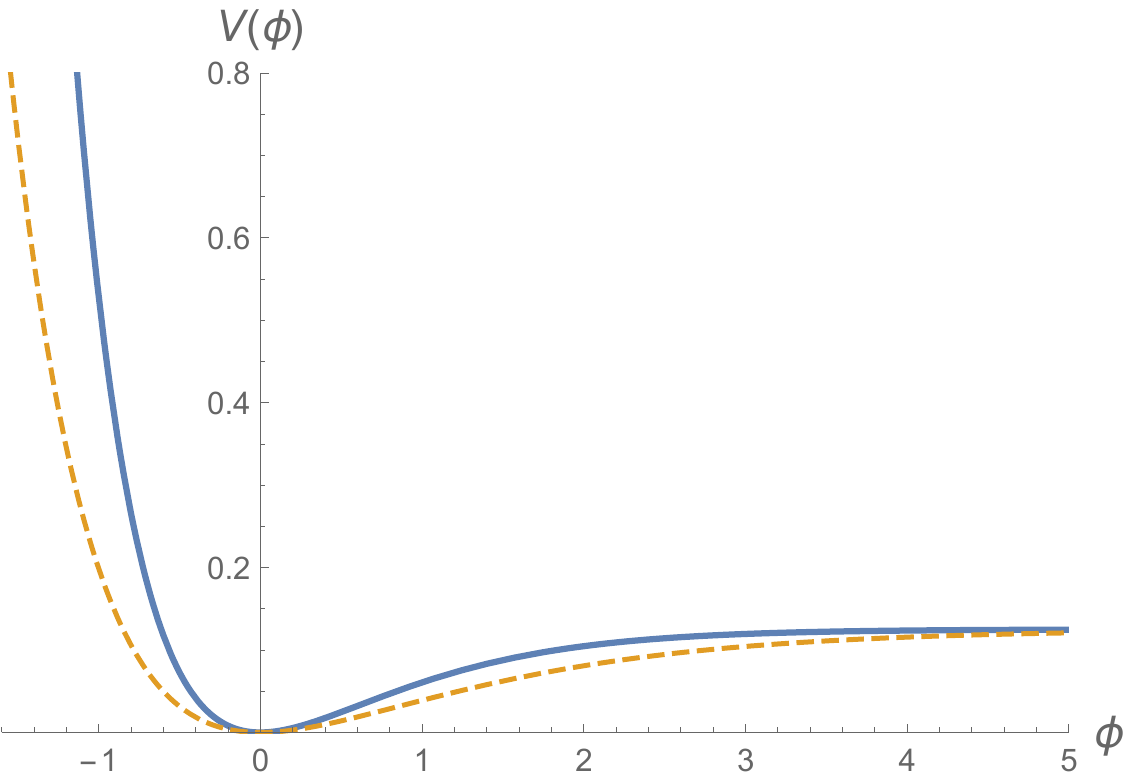}
\caption{\label{fig:06} The Einstein-frame scalaron potential, $V(\phi)$, 
evaluated at $\Lambda = 0$ for our exact square-root deformation 
(solid blue) and the original Starobinsky model (dashed orange). At large 
positive $\phi$, both potentials asymptote to the inflationary plateau at 
$1/(8\alpha)$. However, the deformed potential converges more rapidly 
($\propto \phi e^{-2\sqrt{2/3}\phi}$ versus the standard 
$\propto e^{-\sqrt{2/3}\phi}$), driving a stronger suppression of the 
tensor-to-scalar ratio $r$. As $\phi \to -\infty$, both potentials construct 
a steep energetic barrier. The physical distinction lies in the Jordan-frame 
origin of these walls: the standard Starobinsky wall corresponds to a singular 
truncation of the manifold at a finite curvature $R = -1/(2\alpha)$, beyond 
which the mapping yields unphysical fields. In contrast, the deformed model's 
wall corresponds to the true geometric infinity $R \to -\infty$, ensuring the 
global curvature domain $R \in (-\infty, +\infty)$ is non-singularly mapped 
onto the canonical scalar field space.}
\end{figure}

Using the exact inverse mapping from Eq.~(\ref{eq:Rphi}), we arrive at an 
exact, closed-form expression for the Einstein-frame potential,
\begin{equation}
\label{eq:Vexact}
V(\phi) = \frac{1}{8\alpha} \left[ 1 - \left(1 + 2\sqrt{\frac{2}{3}}\phi\right) \exp\left(-2\sqrt{\frac{2}{3}}\phi\right) \right] 
+ \Lambda \exp\left(-2\sqrt{\frac{2}{3}}\phi\right).
\end{equation}
This can be directly compared to the standard Starobinsky case 
(shown in Fig.\ \ref{fig:06}),
\begin{equation}
\label{eq:VexactStarobinsky}
V_{\text{Starobinsky}}(\phi) = \frac{1}{8\alpha} \left[ 1 - \exp\left(-\sqrt{\frac{2}{3}}\phi\right) \right]^2 
+ \Lambda \exp\left(-2\sqrt{\frac{2}{3}}\phi\right).
\end{equation}
We now analyze the behavior of this analytic potential at the 
theoretical boundaries.

\subsection{The High-Energy Inflationary Plateau}

For large positive values of the scalar field ($\phi \to +\infty$), which 
maps to the early-universe high-curvature regime ($R \to +\infty$), 
the exponential term $e^{-2\sqrt{2/3}\phi}$ rapidly decays to zero. 
The potential therefore strongly asymptotes to a constant,
\begin{equation}
V(\phi \to +\infty) \simeq \frac{1}{8\alpha},
\end{equation}
perfectly reproducing the $\mathcal{O}(1/\alpha)$ flat inflationary 
plateau characteristic of standard Starobinsky gravity. The potential 
naturally supports slow-roll inflation, retaining the phenomenological 
successes of the $R^2$ model regarding the generation of 
scale-invariant primordial perturbations.

\subsection{The Asymptotic Potential Wall and Cosmological Bounce}
\label{sec:theWall}

The true departure from standard $f(R)$ gravity occurs in the negative 
field limit ($\phi \to -\infty$), which maps to the deep negative curvature 
regime ($R \to -\infty$). 

In the standard Starobinsky model, the conformal mapping becomes 
singular at $R_{\text{crit}} = -1/(2\alpha)$, where $f'(R) = 0$. Although 
the Einstein-frame potential diverges to $+\infty$ as the scalar field 
approaches $\phi \to -\infty$, this asymptotic boundary corresponds 
to a \emph{finite} curvature in the Jordan frame, abruptly truncating 
the physical manifold. In the deformed model, this previously critical 
curvature $R_{\text{crit}}$ is smoothly traversed, corresponding to 
a perfectly regular, finite scalar field value,
\begin{equation}
\phi_{\text{crit}} 
= 
\sqrt{\frac{3}{2}}\mathrm{arcsinh}\left(-\frac{1}{2}\right) \approx -0.5894.
\end{equation}

For large negative $\phi$ (corresponding to $R \to -\infty$), our potential, Eq.~(\ref{eq:Vexact}), is
\begin{equation}
V(\phi \to -\infty) 
\simeq 
\frac{1}{8\alpha} \left( 2\sqrt{\frac{2}{3}} |\phi| \right) \exp\left(2\sqrt{\frac{2}{3}}|\phi|\right).
\end{equation}
Because the exponent is strictly positive, the potential increases 
without bound,
\begin{equation}
\lim_{\phi \to -\infty} V(\phi) = +\infty.
\end{equation}
Thus, when the scalar field rolls down the inflationary plateau, crosses 
the minimum at $\phi=0$, and enters the negative curvature regime 
($\phi < 0$), it encounters an exponentially steepening potential barrier. 
Classical dynamics dictates that the field's kinetic energy will be completely 
depleted against this wall. The field is physically forbidden from reaching 
$\phi \to -\infty$ (the $R \to -\infty$ limit), forcing a deterministic, smooth 
bounce. The infinite-gravity strong-coupling singularity is thus entirely 
shielded by an energetic wall.

\section{Slow-Roll Inflation and CMB Observables}

To determine the viability of this deformed $f(R)$ model as a realistic 
theory of the early universe, we analyze its predictions for the primordial 
power spectrum. We work in the Einstein frame, utilizing the exact 
analytical potential derived in Eq.~(\ref{eq:Vexact}).

During inflation, the scalar field is displaced to large positive values 
($\phi \gg 1$), slowly rolling down the potential plateau toward the 
true vacuum at $\phi = 0$. For convenience, we define the characteristic 
mass scale $V_0 = 1/(8\alpha)$ and the dimensionless coupling constant 
$\beta = 2\sqrt{2/3} = \sqrt{8/3}$. The potential in the inflationary regime 
is thus described by
\begin{equation}
V(\phi) = V_0 \left[ 1 - \left(1 + \beta\phi\right) e^{-\beta\phi} \right].
\end{equation}

\subsection{Slow-Roll Parameters}

The dynamics of slow-roll inflation is characterized by the standard 
parameters $\epsilon$ and $\eta$, defined 
(in units where $M_{\mathrm{Pl}} = 1$) as
\begin{equation}
\epsilon = \frac{1}{2} \left( \frac{V'(\phi)}{V(\phi)} \right)^2, 
\quad 
\eta = \frac{V''(\phi)}{V(\phi)}.
\end{equation}
The derivatives of the exact potential are
\begin{eqnarray}
V'(\phi) = V_0 \beta^2 \phi e^{-\beta\phi} ,
\quad
V''(\phi) = V_0 \beta^2 \left( 1 - \beta\phi \right) e^{-\beta\phi}.
\end{eqnarray}
In the deep slow-roll regime where $\phi \gg 1$ and 
$e^{-\beta\phi} \ll 1$, the potential is dominated by the plateau, 
$V(\phi) \simeq V_0$. The slow-roll parameters are thus well 
approximated by
\begin{equation}
\label{eq:epsilon_eta_approx}
\epsilon \simeq \frac{1}{2} \beta^4 \phi^2 e^{-2\beta\phi},
\quad
\eta \simeq -\beta^3 \phi e^{-\beta\phi}.
\end{equation}
The inclusion of the $\phi$ prefactor in the exponential decay 
(a direct consequence of the square-root geometric deformation) 
creates a distinct dynamical scaling compared to standard $R^2$ 
gravity.

\subsection{Exact Number of e-folds and Asymptotic Limits}

To evaluate the cosmological observables, we calculate the number 
of e-folds $N$ from the moment a cosmologically relevant scale exits 
the Hubble horizon (at field value $\phi_*$) to the end of inflation 
(at field value $\phi_{\mathrm{end}}$). In the slow-roll approximation, 
this is given by the integral,
\begin{equation}
N \simeq \int_{\phi_{\mathrm{end}}}^{\phi_*} \frac{V(\phi)}{V'(\phi)} d\phi .
\end{equation}
Substituting the exact Einstein-frame potential 
$V(\phi) = V_0 [ 1 - (1+\beta\phi)e^{-\beta\phi} ]$ and its derivative 
$V'(\phi) = V_0 \beta^2 \phi e^{-\beta\phi}$, the integrand can be 
analytically separated into three distinct terms,
\begin{equation}
\frac{V(\phi)}{V'(\phi)} 
= \frac{1 - (1+\beta\phi)e^{-\beta\phi}}{\beta^2 \phi e^{-\beta\phi}} 
= \frac{e^{\beta\phi}}{\beta^2 \phi} - \frac{1}{\beta^2 \phi} - \frac{1}{\beta} .
\end{equation}
This separation allows for exact integration. The first term is expressed via
the Exponential Integral function, 
\BEq
\mathrm{Ei}(x) \equiv \int_{-\infty}^{x} \frac{e^t}{t} dt, 
\EEq
while the second and third terms yield simple logarithmic and linear 
contributions, respectively. Integrating exactly over the interval yields,
\begin{equation}
N 
= 
\left[
\frac{1}{\beta^2} \mathrm{Ei}(\beta\phi) 
- \frac{1}{\beta^2} \ln(\phi) - \frac{\phi}{\beta} 
\right]_{\phi_{\mathrm{end}}}^{\phi_*}.
\end{equation}

To extract physical predictions, we evaluate this expression in the 
standard cosmological limits. Because the field value at horizon 
exit is significantly larger than at the end of inflation 
($\phi_* \gg \phi_{\mathrm{end}}$), the evaluation at the upper 
limit via the rapidly growing Exponential Integral overwhelmingly 
dominates. The contribution from the lower limit 
$\phi_{\mathrm{end}}$ is $\mathcal{O}(1)$, which is negligible 
compared to the $N \simeq 60$ e-folds generated on the plateau.

Applying the asymptotic expansion for the Exponential Integral 
at large arguments ($\beta\phi_* \gg 1$),
\begin{equation}
\mathrm{Ei}(\beta\phi_*)
     \simeq
      \frac{e^{\beta\phi_*}}{\beta\phi_*} 
      \left( 1 + \frac{1}{\beta\phi_*} + \frac{2}{(\beta\phi_*)^2} + \dots  \right),
\end{equation}
and taking the leading-order term, the expression for the total 
number of e-folds simplifies to the asymptotic approximation,
\begin{equation}
\label{eq:N_approx}
    N \simeq \frac{e^{\beta\phi_*}}{\beta^3 \phi_*}.
\end{equation}
This algebraic relationship allows us to invert $N(\phi_*)$ to express 
the slow-roll parameters directly in terms of the observable number 
of e-folds. Substituting Eq.~(\ref{eq:N_approx}) into the asymptotic 
forms of the slow-roll parameters, and recalling that $\beta = \sqrt{8/3}$, 
we obtain,
\begin{equation}
\epsilon \simeq \frac{3}{16 N^2}, \quad \eta \simeq -\frac{1}{N}.
\end{equation}

\subsection{Comparison with Observational Data}

We can now extract the definitive cosmological predictions of the 
model. To leading order in slow-roll, the scalar spectral index $n_s$ 
and the tensor-to-scalar ratio $r$ are given by
\begin{equation}
    n_s \simeq 1 - 6\epsilon + 2\eta \simeq 1 - \frac{2}{N},
    \quad
    r \simeq 16\epsilon \simeq \frac{3}{N^2}.
\end{equation}
The predicted spectral index $n_s$ is identical to the universal attractor 
prediction of the standard Starobinsky model. For a nominal value of 
$N = 60$ e-folds, this yields $n_s \simeq 0.967$, placing the model 
within the Planck and BICEP/Keck array cosmological 
constraints \cite{Planck2018, BICEPKeck2021}.

Importantly, the tensor-to-scalar ratio, $r \simeq 3/N^2$, presents 
a measurable phenomenological divergence from standard $R^2$ 
gravity (which predicts $r \simeq 12/N^2$). The proposed deformation 
suppresses the generation of primordial gravitational waves by 
a factor of 4. For $N=60$, we predict $r \simeq 0.00083$. This deeper 
suppression satisfies the increasingly stringent upper bounds on $r$ 
placed by recent CMB polarization observations and provides a distinct, 
falsifiable signature for future high-precision CMB observatories.

In summary, the analytical structure required to resolve the 
strong-coupling singularity not only preserves the theory's inflationary 
viability, but also yields predictions comfortably within the observationally 
favored parameter space.

\section{Discussion and Conclusion}

The Starobinsky $R^2$ model is one of the most observationally 
successful implementations of modified gravity, providing a natural 
mechanism for early-universe inflation without the need to introduce 
additional scalar fields. However, its strict reliance on 
$f'(R) = 1 + 2\alpha R$ limits its domain of validity. The existence 
of a critical negative curvature where the effective gravitational 
coupling diverges ($f'(R) = 0$) and the scalaron mass becomes 
ill-defined presents a severe physical singularity, preventing 
the theory from consistently describing collapsing spacetimes 
or bouncing cosmologies.

In this paper, we have introduced a geometric deformation of 
the Starobinsky model that resolves this singularity. By defining 
the derivative of the action through a hyperbolic square-root 
structure, $f'(R) = \alpha R + \sqrt{\alpha^2 R^2 + 1}$, we construct 
a globally stable theory valid across the entire curvature domain 
$R \in (-\infty, +\infty)$. This model recovers general relativity in 
the low-curvature limit, preserves the $\mathcal{O}(R^2)$ inflationary 
dynamics at large positive curvatures, and mathematically forbids 
the derivative of the action from crossing zero.

A notable feature of the exact action, Eq.~(\ref{eq:fR}), is the presence 
of the $\mathrm{arcsinh}(\alpha R)$ term. Using the standard identity 
for the inverse hyperbolic sine, this term can be rewritten in an explicit 
logarithmic form, establishing a connection between this geometric 
deformation and the broader class of logarithmic $f(R)$ gravity theories 
(e.g., \cite{Sadeghi2015, Amin2016}). Logarithmic modifications to the 
Einstein-Hilbert action frequently appear in the literature, often originating 
from quantum-gravitational loop corrections (which typically generate 
terms of the form $R^2 \ln(R/\mu^2)$) or phenomenological models 
addressing late-time dark energy. 

However, standard logarithmic theories inherently suffer from 
domain restrictions: the argument of a simple $\ln(R)$ term becomes 
complex or undefined for $R < 0$, rendering the theory non-viable 
in negative-curvature spacetimes. The proposed model circumvents 
this limitation. Because the argument of our logarithmic formulation 
is $\alpha R + \sqrt{\alpha^2 R^2 + 1}$, it remains strictly positive 
for all real values of $R$. The logarithm is therefore globally real, 
smooth, and well-defined, allowing the theory to transition from 
positive to negative scalar curvatures without introducing imaginary 
action components or branch-cut discontinuities.
Beyond logarithmic approaches, other recent theoretical efforts have 
employed hyperbolic functions to regularize $R^2$ gravity. For instance, 
$\tanh$-based Lagrangians have been investigated by Bazeia and Lima 
\cite{Bazeia2024} to support localized structures and well-behaved scalar 
field profiles within the Starobinsky framework. 

The proposed hyperbolic square-root ansatz offers a distinct advantage for global spacetime 
evolution: it provides an exact, unbounded, and strictly monotonic mapping 
that preserves the $f''(R) > 0$ stability condition for all $R$, avoiding any 
artificial saturation of the curvature corrections at high energies.
When mapped to the Einstein frame, this unbroken global mapping yields 
a direct physical consequence. Rather than terminating at a finite geometric 
cusp, the scalar potential $V(\phi)$ develops a steep, unbounded energetic 
barrier as $\phi \to -\infty$. This structure physically censors the 
infinite-curvature singularity by reflecting the scalar field, offering a consistent 
mechanism for a non-singular bouncing cosmology. 

Importantly, enforcing this mathematical consistency does not degrade 
the model's predictive power for the early universe. As shown, the resulting 
scalar potential alters the standard inflationary predictions by introducing 
a unique scaling in the slow-roll parameters, yielding a tensor-to-scalar ratio 
of $r \simeq 3/N^2$. This provides a stronger suppression of primordial 
gravitational waves than the standard Starobinsky model, placing the theory 
comfortably within the parameter space favored by the baseline CMB polarization 
data \cite{Planck2018, BICEPKeck2021}.

Finally, it is worth commenting on the phenomenological outlook of 
the model in light of recent observational trends \cite{Odintsov2026}. 
While the hyperbolic 
square-root deformation successfully cures the classical strong-coupling 
singularities of the Starobinsky model and maintains a globally stable 
inflationary plateau, it naturally inherits the standard predictions of 
$R^2$-like attractors, namely $n_s \simeq 1 - 2/N$ and a strongly 
suppressed $r$. Recent data combinations from high-resolution CMB 
surveys and BAO measurements \cite{ACTDR6, DESI2024, McDonough2025} 
have shown an emerging preference for a slightly higher 
scalar spectral index, $n_s \gtrsim 0.97$. Accommodating 
this upward shift within the strict $1 - 2/N$ scaling would require a larger 
number of $e$-folds ($N \gtrsim 70$), implying a non-standard 
post-inflationary cosmological history. We leave the exploration 
of such phenomenological extensions, as well as the detailed reheating 
kinematics of our globally regularized Einstein-frame potential, to future study.

\end{document}